\documentclass[conference]{IEEEtran}
\IEEEoverridecommandlockouts
\usepackage{cite}
\usepackage{amsmath,amssymb,amsfonts}
\usepackage{algorithmic}
\usepackage{graphicx}
\usepackage{textcomp}
\usepackage{xcolor}
\def\BibTeX{{\rm B\kern-.05em{\sc i\kern-.025em b}\kern-.08em
    T\kern-.1667em\lower.7ex\hbox{E}\kern-.125emX}}
\begin{document}

\title{Layering the monitoring action for improved flexibility and overhead control: work-in-progress
}

\author{\IEEEauthorblockN{Giacomo Valente\textsuperscript{1}, Tiziana Fanni\textsuperscript{2}, Carlo Sau\textsuperscript{3}, Francesco Di Battista\textsuperscript{1}}
\IEEEauthorblockA{\textsuperscript{1}Università degli Studi dell'Aquila, \textsuperscript{2}Università degli Studi di Sassari, \textsuperscript{3}Università degli Studi di Cagliari \\
\texttt{email address: giacomo.valente@univaq.it, tfanni@uniss.it, carlo.sau@diee.unica.it}}
}

\maketitle

\begin{abstract}
 With the diffusion of complex heterogeneous platforms and their need of characterization, monitoring the system gained increasing interest. This work proposes a framework to build custom and modular monitoring systems, flexible enough to face the heterogeneity of modern platforms, offering a predictable HW/SW impact. 
\end{abstract}

\begin{IEEEkeywords}
SW monitoring, HW Monitoring, self-awareness, monitoring layer
\end{IEEEkeywords}

\section{Context and Objectives}

Nowadays, embedded platforms are evolving toward heterogeneous architectures, including general and specific purpose processors implemented on chips with both 
dedicated and reconfigurable logic. 
This evolution has been mainly driven by the need of 
different functionalities while trading-off among non-functional requirements (e.g., timing, energy, cost) \cite{2020_Valente, sau_2017}. 
In this context, the demand of system characterization techniques is increasing: simulation not always represents an acceptable solution, since a fine granularity requires complex models and tends to slowdown the application (with respect to its actual running time \cite{Doyle_2017}), qualifying the usage of runtime monitoring systems \cite{Kornaros_2013}. 
However, the evolution toward heterogeneous architectures also impacts on monitor design, since it (i) elicits fresh \textit{monitoring ReQuireMents} (RQMs), and (ii) requires the satisfaction of traditional RQMs (e.g., monitor of power dissipation, monitor of execution time) on those new platforms. 
As a result, there are consolidated solutions dedicated to specific areas, targeting specific platform components and RQMs. The drawback of this approach is highlighted when we try their adaptation for monitoring new platforms (\textit{flexibility}), and also on evolving them toward satisfaction of fresh RQMs (\textit{applicability}). Furthermore, the need to satisfy multiple RQMs leads to the adoption of multiple different monitoring systems, with a difficult to predict HW/SW impact on system resources (\textit{predictability}).
Many academic and industrial solutions are available in literature to address these issues.
The works in \cite{Seo_2018}, \cite{Valente_2016}, \cite{Valente_2015},
\cite{Doyle_2017}, and
\cite{Fanni_2019} all offer a solution to compose a monitoring system in a modular way, starting from the basic events to be monitored and building the monitor to capture and store them.
Works in \cite{Goeders_2017} and \cite{Hammouda_2017} both allow to design and introduce a monitoring system during a \textit{High-Level Synthesis} (HLS) flow.
Finally, \cite{Najem_2017} and \cite{Zoni_2018} start from simulation of the system, trying to understand which are the target signals to be monitored against some RQMs, by exploiting data-mining techniques.
However, both the approaches based on HLS and data-mining provide monitors difficult to combine with SW-tasks, due to a monitor creation that does not take into account microprocessors architectures, limiting their flexibility. Modular monitoring solutions represent a promising approach, but the available ones lack in applicability to different RQMs, being typically focused on specific purposes, i.e., debugging \cite{Seo_2018}, or timing performance \cite{Valente_2016,Doyle_2017}. Only the work in \cite{Fanni_2019}  targets both performance- and debugging-oriented monitoring over heterogeneous architectures. 
By looking at industrial solutions, two modular approaches can be mentioned: ARM Coresight \cite{ARM_2013} and AXI Performance monitor \cite{axi_mon}. 
The former is not flexible enough to be used on custom accelerators, while the latter only targets bus interconnections.
The main contribution of this work in progress paper is a \textit{HW monitoring layer}, part of a larger project that aims at providing a framework for building custom hardware monitoring systems, that, differently from the state of art approaches, offers flexibility, applicability, and predictability for the generated monitors.The HW monitoring layer is the component that allows to provide those features by allowing the build of the monitors in a modular way.
The final framework will target heterogeneous architectures with reconfigurable accelerators implemented on FPGA, taking as input a model of the platform description and RQMs, producing an HDL description of the monitored platform as output. The resulting monitored platform will be provided with monitors satisfying the given RQMs, with a well defined monitoring overhead in terms of \textit{area}, \textit{PoWeR} (PWR), and \textit{SoftWare OVerhead} (SWOV). 
The proposed framework is similar to \cite{Valente_2016} and \cite{Seo_2018} but additionally, besides targeting hardware accelerators (that could be addressed with an extensions of such previous works), it allows optimal resource utilization leveraging on event instances sharing.

\section{The HW Monitoring Layer}

\begin{table*} [!h]
\caption {Test Results. Int., DataM and Core refer respectively to event triggers placed, sticking to the proposed HW layer, in Interconnection, Data Manager and Core (1E32 is 1 EVMON 32-bit size, 1T64 is 1 TMON 64-bit size, P is programmability). A term of comparison for LUT and FF is provided: in this case \cite{axi_mon} is used to monitor Int. and \cite{xilinx_ila} to monitor DataM and Core. }
\label{tab:exper_results}
\centering
\begin{tabular}{ |c|c|ccc||c|c|c|c||c|c| }
\hline
ID & RQM & Int. & DataM & Core & LUT & FF & PWR[mW] & SWOV[us] & LUT \cite{axi_mon} \cite{xilinx_ila} & FF \cite{axi_mon} \cite{xilinx_ila} \\
\hline
Y0 & - & - & - & - & 3397 & 2864 & 24 & 10.981 & - & - \\
\hline
Y1 & RQM1+RQM2+RQM3 & (1E32)(P) & (1T64) & (2E10) & +9.45\% & +13.2\%  & +8.33\%  & +39.68\% & +82.16\% & +166.3\%  \\
\hline
Y2 & RQM2+RQM3+RQM4 & - & (1T64) & (2E10) & +3.39\% & +8.66\%  & +8.33\%  & +37.33\% & +68.11\% & +40.12\%  \\
\hline
Y3 & RQM1+RQM4+RQM5 & (1E32)(P) & (1T64) & - & +8.98\% & +12.33\%  & +12.5\%  & +37.36\% & +82.16\% & +166.3\%  \\
\hline
\end{tabular}
\end{table*}

\begin{figure}[h!t]
  \centering
  \includegraphics[width=1\linewidth,trim={0 0.8cm 0 0.4cm}]{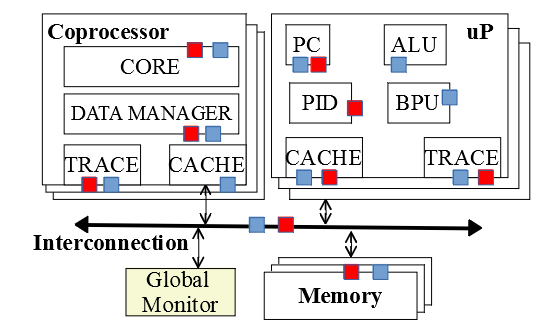}
  \caption{Heterogeneous reference platform   with places for performance event triggers (blue) and for debug ones (red).  }
  \label{fig_monitor_performance}
\end{figure}

As starting point for the construction of the HW monitoring layer, we considered the generic online monitoring process proposed by Kornaros et al. \cite{Kornaros_2013}. Authors state that a monitoring process has five phases: event trigger, data capture, filtering, decision, and reaction.
To this end, we divided the construction in two main phases: Ph.1) identification of places for event triggers to have a monitoring action that is both applicable for satisfying fresh RQMs and flexible; Ph.2) building of a HW layer to implement the monitoring action.
In Ph.1 RQMs are covered in a general way (thus enabling applicability). We considered the six classes of RQMs for a monitoring action 
\cite{Kornaros_2013}: \textit{Monitor for DeBuG} (MDBG), \textit{PerFormance} (MPF),  \textit{Power/Energy/Temperature} (PET), \textit{Quality of Service} (QoS), \textit{Fault Tolerance/Reliability} (FT), and \textit{Security} (Sec).
Then, to guarantee the flexibility, we associated those RQMs to a general reference platform for embedded systems, identifying the places for event triggers (see Fig. \ref{fig_monitor_performance}). Analysing the output of Ph.1, we noticed that multiple RQMs can share the same trigger location.

\begin{figure}[h!t]
  \centering
  \includegraphics[width=1\linewidth,trim={0 0.8cm 0 0.4cm}]{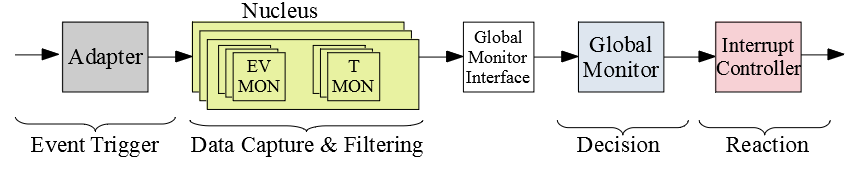}
  \caption{The proposed hardware layer.}
  \label{fig_monitor_actors}
\end{figure}

Sharing the event triggers among multiple RQMs allows to share, in turn, the collected events: in the proposed HW layer, built in Ph.2 (see Fig. \ref{fig_monitor_actors}), an \textit{adapter} block samples the i-th event instance and sends it to the data capture and filtering blocks. For these last two phases, a customizable number of \textit{nucleus} blocks is provided to selectively capture different data according to RQMs (enabling applicability). Each nucleus can aggregate events, in form of event instances coming from adapter, by means of multiple \textit{event monitors} (EVMON) and \textit{time monitors} (TMON). The aggregation basically reflects the function mapping events to metrics (e.g. \emph{start} and \emph{done} signals assertion to measure an execution time). 
Nucleus data are then sent to a \textit{global monitor interface} (GMI), that sends the information toward a \textit{global monitor} (GM), connected at the same level of the reference platform actors; GM implements decision making stage and triggers a reaction by means of an \textit{interrupt controller}. In case of target architecture change, only adapter and nucleus need to be modified, enabling flexibility.
To illustrate the benefits of the proposed solution in generating monitoring systems, we defined some RQMs for an embedded application executed on a heterogeneous platform implemented, using Xilinx Vivado 2017.4, on a Zynq7000 XC7Z020 \cite{xilinx_zynq}. 
It has an AXI system bus connecting a dual-core ARM processor, an external DRAM memory and a hardware accelerator generated with the MDC suite\footnote{Multi-Dataflow Composer: 
https://github.com/mdc-suite/mdc
} \cite{Sau_2016}.
The application, running on the ARM processor, prepares some inputs in the external DRAM and triggers a DMA to transfer them to the accelerator. This latter performs multiply-and-accumulate operations (constituting the HW task) and stores back the result in the DRAM through a new DMA mediated transfer.
The considered RQMs are the followings: RQM1 (MDBG - data transfer fault detection on the accelerator), RQM2 (MPF - execution time of the HW task), RQM3 (MDBG - accelerator computation fault detection), RQM4 (MDBG - watchdog for the HW task), RQM5 (MPF - throughput of data processed by HW task). 
Table \ref{tab:exper_results} reports area, PWR (only dynamic FPGA fabric one), and SWOV impact of the different monitoring solutions depending on the considered input RQMs (Y1, Y2, and Y3). A comparison, only in terms of resources, with commercial monitoring solutions (\cite{axi_mon} and \cite{xilinx_ila}) shows the better performance of the proposed solution in satisfying different RQMs. 

\section*{Acknowledgment}
This work is part of the FitOptiVis project \cite{zaid_2019}, funded by the ECSEL Joint Undertaking under grant number H2020-ECSEL-2017-2-783162, and of the  Comp4Drones project No. 826610, ECSEL-JU 2018.


\end{document}